%% file: paper4.tex
\documentclass[
	amsmath,
	amssymb,
    aps,
    reprint,
    pre,
    showkeys,
]{revtex4-1}

\usepackage[utf8]{inputenc}
\usepackage{amsthm}
\usepackage{hyperref}
\usepackage{graphicx}
\usepackage[font=small,justification=centerlast,format=plain]{caption}
\usepackage{subcaption}
\usepackage{xcolor}
\usepackage{dsfont}
\usepackage{booktabs}
\usepackage{bm}
\usepackage[capitalise]{cleveref}

\begin{document}

\title{Multi-body Interactions and Non-Linear Consensus Dynamics on Networked Systems}
\author{Leonie Neuhäuser}
\email{neuhauser@maths.ox.ac.uk}
\affiliation{Mathematical Institute, University of Oxford, Oxford, UK}
\affiliation{Hertie School, Berlin, Germany}
\author{Andrew Mellor}
\email{mellor@maths.ox.ac.uk}
\affiliation{Mathematical Institute, University of Oxford, Oxford, UK}
\author{Renaud Lambiotte}
\email{renaud.lambiotte@maths.ox.ac.uk}
\affiliation{Mathematical Institute, University of Oxford, Oxford, UK}

\date{\today}

\newtheorem{theorem}{Theorem}[section]
\newtheorem{proposition}[theorem]{Proposition}
\newtheorem{corollary}[theorem]{Corollary}
\newtheorem{lemma}[theorem]{Lemma}
\theoremstyle{definition}
\newtheorem{definition}[theorem]{Definition}
\newtheorem{assumption}{Assumption}
\theoremstyle{remark}
\newtheorem*{remark}{Remark}

\renewcommand\theparagraph{(\alph{paragraph})}

\begin{abstract}
	Multi-body interactions can reveal higher-order dynamical effects that are not captured by traditional two-body network models.
	In this work, we derive and analyse  models for consensus dynamics on hypergraphs, where nodes interact in groups rather than in pairs.
	Our work reveals that  multi-body dynamical effects that go beyond rescaled pairwise interactions can only appear if the interaction function is non-linear, regardless of the underlying multi-body structure.
	As a practical application, we introduce a specific non-linear function to model three-body consensus, which incorporates reinforcing group effects such as peer pressure.
	Unlike consensus processes on networks, we find that the resulting dynamics can cause shifts away from the average system state.
	The nature of these shifts depends on a complex interplay between the distribution of the initial states, the underlying structure and the form of the interaction function.
	By considering  modular hypergraphs, we discover state-dependent, asymmetric dynamics between polarised  clusters where multi-body interactions make one cluster  dominate the other.
\end{abstract}


\maketitle

\section{Introduction}
\label{sec:introduction}

In recent years, researchers have developed a variety of higher-order models for complex systems in order to enrich the standard network formalism when it is not sufficient to capture their structure and function \cite{lambiotte_networks_2019}. 
Among those, multi-body models, sometimes called combinatorial models, focus on the importance of group interactions, that is situations when the basic unit of interaction involves more than two nodes. Multi-body interactions are observed in different areas in nature \cite{santos_topological_2018}, society \cite{iacopini_simplicial_2019} and technology \cite{olfati-saber_consensus_2007}, with examples ranging from collaborations of authors \cite{patania_shape_2017} to neuronal activity \cite{giusti_clique_2015,reimann_cliques_2017}. Such systems may be represented as hypergraphs or simplicial complexes, and 
a substantial body of work has characterised their structural properties. However, a proper understanding of how multi-body interactions affect spreading dynamics in networked systems is still nascent \cite{bick2016chaos,schaub_random_2018,iacopini_simplicial_2019,arruda2019social, skardal2019abrupt}.

When considering a dynamical process driven by multi-body interactions,  it is important to distinguish between an accumulation of two-body dynamics and genuine multi-body dynamics. In the former case, the influence on an agent can be fully explained by its pairwise relationships to other group members. In contrast, multi-body dynamical systems additionally account for the effect of the group as a whole. An area where this distinction is often blurred is the modelling of complex diffusion  in social networks.
The adoption of norms and opinion between agents is known to be a non-trivial process, which may drastically differ from a simple model of epidemic spreading. Experiments in social psychology such as the conformity experiment \cite{asch_effects_1951} indicate that multiple exposures might be necessary for an agent to adopt a certain state. This phenomenon may be described with threshold models, in which the state of agents switches if a certain fraction of their neighbours agrees. Observe that these models are based on independent, two-body interactions that are linearly accumulated and therefore do not account for multi-body effects resulting from those edges being part of a closed group \cite{chang_co_diff_2018}. 
Yet, it is well known in sociology that the dynamics in a social clique are determined not just by the pairwise relationships of its members, but often by complex mechanisms of peer influence and reinforcement \cite{petri_simplicial_2018, milgram1969note, latane1981psychology}.

There are different ways of encoding  the multi-body structure of a networked system. They range from set systems and general hypergraphs \cite{noauthor_book_1987,arruda2019social} to approaches adapted from algebraic topology based on simplicial complexes \cite{schaub_random_2018,mukherjee_random_2016,parzanchevski_simplicial_2017,muhammad_control_nodate,petri_simplicial_2018, masulli_algebro-topological_nodate,salnikov2018simplicial}. 
Hypergraphs generalise standard networks by allowing nodes to interact through hyperedges of different sizes. Size two corresponds to two-body edges, size three to three-body edges and, in general, size $k$ to $k$-body edges, associated to $k$-cliques.
Importantly, different types of triangles may be found in hypergraphs, those obtained by combining three two-body edges, and those corresponding to one three-body edge, and they have a different nature.
In simplicial complex approaches, a fully connected group of $d+1$ nodes is formalised as a $d$-simplex. The unity of all simplices that are formed by the underlying edge structure of a graph is called a simplicial complex and can be used as the underlying topology \cite{petri_simplicial_2018,schaub_random_2018}. These complexes can be broken down into different simplex skeletons that depend on the dimension of simplices they include. The 1-skeleton of a simplicial complex just includes (two-body) edges and is equivalent to the conventional network structure. The 2-skeleton includes all simplices up to dimension two (nodes, edges and triangles) and a general $d$-skeleton all simplices up to dimension $d$ \cite{schaub_random_2018,parzanchevski_simplicial_2017}.

Hypergraphs and simplicial complexes are natural languages to describe the structure of multi-body networked systems. 
However, understanding how these substrates affect the dynamics requires an additional modelling step. 
Multi-body  dynamical models that use simplicial complexes may, for instance, exploit an algebraic structure called the Hodge-Laplacian, analogous to the graph Laplacian \cite{schaub_random_2018}.
However, the resulting diffusive process becomes more complicated to study, as diffusion is now defined between simplices of any dimension, and not only between nodes as is usually the case on networks.
This generalisation leads to challenges in both analysis and  interpretability,  as state variables can usually be measured only on nodes, and not on other network entities, in  empirical data. In this paper, we explore a different modelling approach based on hypergraphs where, more realistically, the states are defined on the nodes, and their evolution depends on the combined values  
of all the nodes inside each multi-body interacting unit. For the sake of simplicity, we focus mostly on three-body dynamical systems as a first extension to two-body dynamical systems \cite{vasile_sen_2015}, that is we consider the dynamics taking on an hypergraph where all hyperedges have size three.
We show that linear dynamics on hypergraphs can always be rewritten as a dynamics on a standard, two-body network. This observation emphasises that non-linear interactions are necessary for the multi-body dynamical system not to be reducible to a two-body dynamical system. As a next step, we propose and analyse a minimal non-linear model for consensus dynamics including reinforcing group effects, which we study analytically and by means of numerical simulations.

The remainder of this article is outlined as follows.
In \Cref{sec:model} we describe the model for three-body dynamical systems.
We show further in \Cref{sec:interaction_function} the role of the interaction function and differentiate between the linear and non-linear case.
\Cref{sec:higher_order_effects} is devoted to the analysis of a specific non-linear interaction function which captures reinforcement behaviour in triangles.
Finally in \Cref{sec:discussion} we discuss the consequences of incorporating multi-body interactions and propose future research directions.

\section{From two-body to three-body interactions}
\label{sec:model}

Two-body dynamical systems describe general, possibly non-linear, dynamics on conventional edge-based networks \cite{vasile_sen_2015}. Let $\mathcal{G}$ be a network consisting of a set $V(\mathcal{G})=\{1,\dots,N\}$ of $N$ nodes connected by a set of  edges $E(\mathcal{G})=\{(i,j): i,j \in V(\mathcal{G})\}$, described by ordered tuples of nodes. The structure of the network can  be  represented by the adjacency matrix $A \in  \mathbb{R}^{N \times N}$ with entries
\begin{align}
	A_{ij}=\begin{cases}1 & (i,j) \in E(\mathcal{G} ) \\
		0 & \text{ otherwise}.\end{cases}
\end{align}
For undirected networks, which we consider in this paper, $A$ is a symmetric matrix.

Each node $i \in  V (\mathcal{G})$  is endowed with a dynamical variable, $x_i \in \mathbb{R}$ for $i \in \{1, \dots, N\}$, whose evolution is determined by the underlying graph $\mathcal{G}$, defining the edges over which interactions take place, and the set of interaction functions 
\begin{align}
\label{eqn:twoway}
	\mathcal{F}=\left\{f_{i j} | f_{i j} : \mathbb{R}^2 \rightarrow \mathbb{R},(i, j) \in E(G) \right \},
\end{align}
 quantifying how the states of neighbouring nodes affect each other. The time-evolution of the system is then defined by
\begin{align}
\label{eqn:pairwise_interaction}
	S_{i}\left(x_{1}, \ldots, x_{N}\right):=\dot{x}_{i}=\sum_{j =1}^N A_{ij}f_{i j}\left(x_{i}, x_{j}\right),
\end{align}
where the total interaction on a node is the sum, over its neighbours, of possibly non-linear interactions. Particular examples of (\ref{eqn:pairwise_interaction}) include the Kuramoto model  \cite{arenas2008synchronization}, continuous-time random walks and linear consensus on networks \cite{masuda_random_2017}.

For three-body dynamical systems, the topology is encoded by  a hypergraph $\mathcal{H}$  consisting of a set $V(\mathcal{G})=\{1,\dots,N\}$ of $N$ nodes connected by a set of three-body interactions, or triangles.
We describe undirected triangles $T(\mathcal{G})=\{ \{i,j,k\}: i,j,k \in V(\mathcal{G})\}$ as unordered triples of nodes. The structure of the network can be  represented by the adjacency tensor $A \in  \mathbb{R}^{N \times N \times N}$ with entries
\begin{align}
	A_{ijk}=\begin{cases}1 & \mbox{ if } \{i,j,k\} \in T(\mathcal{G} ) \\
		0 & \text{ otherwise}.\end{cases}
\end{align}
The adjacency tensor is symmetric in all dimensions. 
Moreover, we assume that the graph has no self-loops, to avoid classifying edge-combinations involving self-loops as triangles. 
Thus, $A_{ijk}=0$ if $i=j, j=k$, or $i=k$.
By this definition, we consider a system that consists only of three-body interactions, since $A$ cannot describe two-body edges.

\begin{figure}
	\centering
	\begin{subfigure}{0.6\linewidth}
				\hspace{-0.95cm}
		\def\svgwidth{1\textwidth}
		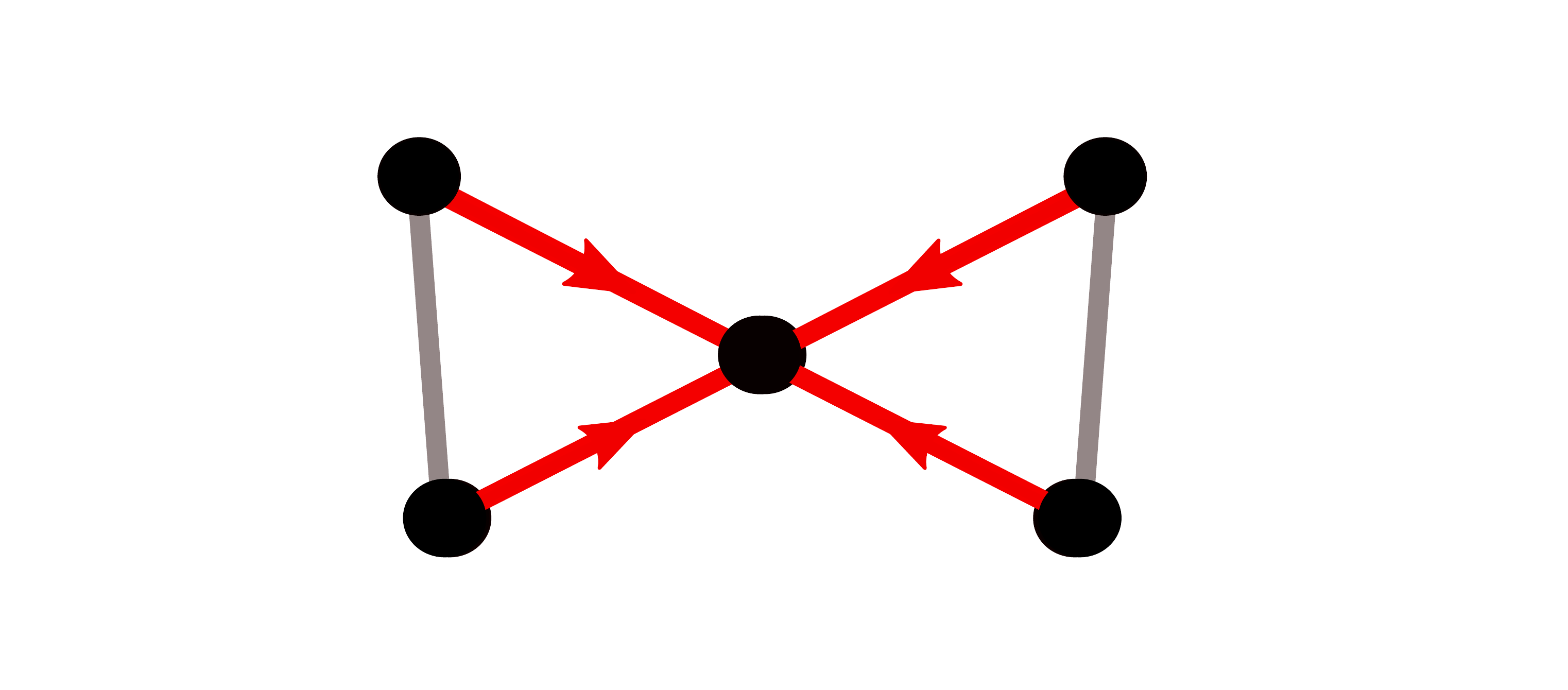
		\caption{Two-body interactions}
	\end{subfigure}
	\begin{subfigure}{0.67\linewidth}
		\def\svgwidth{1\textwidth}
		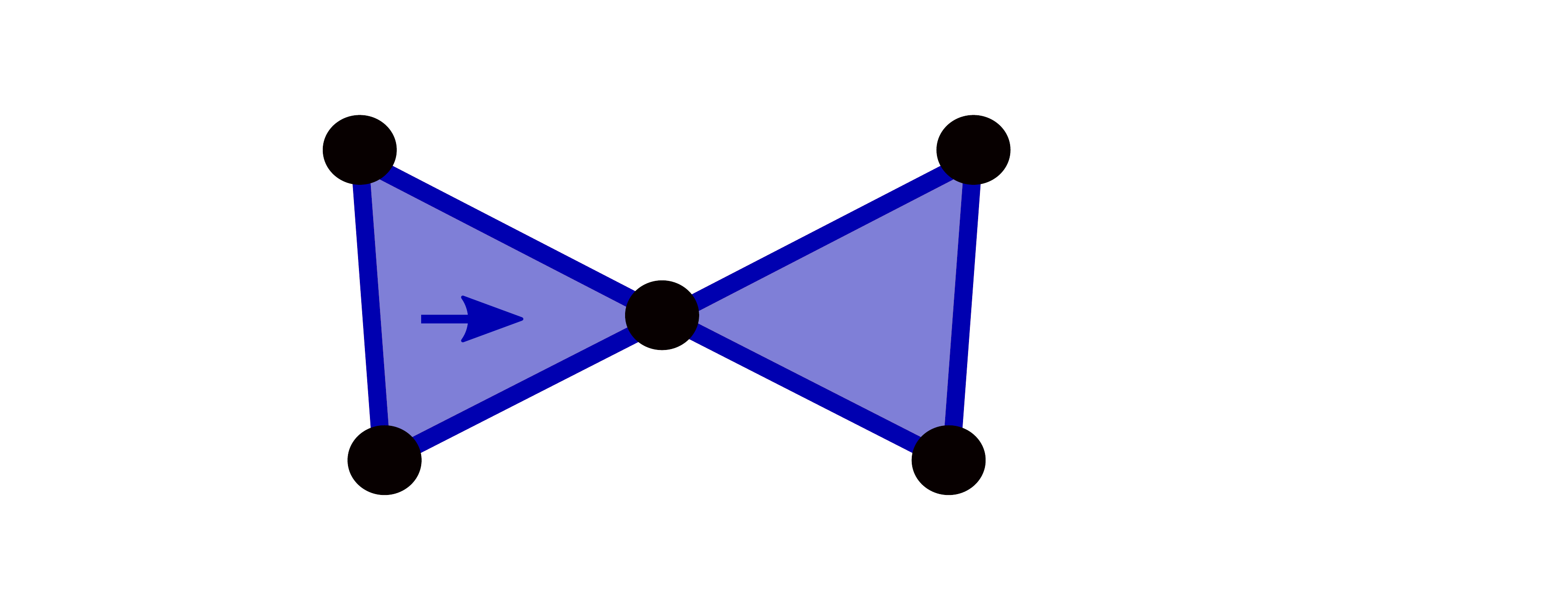
		\caption{Three-body interactions}
	\end{subfigure}
	\caption{The total interaction $S_{i}\left(x_{1}, \ldots, x_{N}\right)$ on node $i$ is the sum of all two-body interactions $f_{ij}$ in (a) and of all three-body interactions $f_i^{\{jk\}}$ in (b).}
	\label{fig:pairwise_threeway_image}
\end{figure}

A three-body dynamical system generalising Eq.~\eqref{eqn:pairwise_interaction} is defined by its structure, encoded  by $\mathcal{H}$, and by the set of interaction functions 
\begin{align}
	F=\left\{f^{(jk) }_i | f_i^{( j k)} : \mathbb{R}^3 \rightarrow \mathbb{R},\{\{i,j,k\}  \in T(G) \right \}
\end{align}
quantifying, for each triangle $\{i,j,k\}$, the combined influence of the states of nodes $j$ and $k$ on the state $i$. The dynamics of the variables $x_i \in \mathbb{R}$ on the hypergraph is then given by
\begin{align}
\label{eqn:threeway}
	S_{i}\left(x_{1}, \ldots, x_{N}\right):=\dot{x}_{i}=\sum_{j , k=1}^N A_{ijk}f^{(jk)}_i\left(x_{i}, x_{j}, x_k\right).
\end{align}
In an undirected setting, it is natural for the symmetries of $f^{(jk)}_i$ to be aligned with the symmetries of $A$.
Therefore, we assume that $f^{(jk)}_i=f^{(kj)}_i :=f_i^{\{jk\}}$ for each nodes. 
The differences between two-body or three-body interactions  are visualised in \Cref{fig:pairwise_threeway_image}.

\section{Non-linearity is necessary for multi-body dynamics}
\label{sec:interaction_function}

\subsection{Symmetries and quasi-linearity}

Our main goal is to examine if, and under which conditions, a three-body dynamical system (\ref{eqn:threeway}) can not be rewritten as a two-body dynamical system (\ref{eqn:twoway}), that is when a multi-body formalism is truly necessary to capture the complexity of a system. We tackle this problem for classes of interaction functions satisfying desirable symmetries. As is often the case for models of non-linear consensus or synchronisation on standard networks, we require  the process to be invariant to translation and rotation.
This is a reasonable assumption for physical and sociological interaction processes ensuring independence on the global reference frame.
A function is rotational and translational invariant if it is invariant under application of elements from the special Euclidean group $\text{SE}(N)$, which is defined as the symmetry group of all translations and rotations around the origin.  As we restrict the scope to scalar values $x_i$ on nodes, and do not consider vectors here, the rotational invariance simply means invariance under a change of signs of the values.
In the case of two-body dynamical systems, it is known that a necessary and sufficient condition for these symmetries to be satisfied is the quasi-linearity of the interaction \cite{vasile_sen_2015}, that is 
\begin{align}
\label{eqn:quasi-linearity}
S_i=\sum_{j}A_{ij}  \, k_{ij}(|x_j-x_i|)  \, (x_j-x_i),
\end{align}
where $k_{ij}$ is an arbitrary function from $\mathbb{R}$ to $\mathbb{R}$. This form implies that the interaction function is, for each edge, an odd function of $(x_j-x_i)$, which is a popular choice in the study of non-linear consensus \cite{Srivastava2011}. Within the language of non-linear consensus, this model belongs to the family of relative non-linear flow. 
Also note that, as long as this function is positive-definite, the dynamics result in a time rescaled  process, which will finally lead, for any initial condition, to a consensus in which all node states have the same value. 
While we cannot generally transfer these results to three-body dynamical systems, they will provide us a guide on how to define a `minimal non-linear' model in that case.
\subsection{Linear Dynamics and Motif Matrices}
\label{sec:linear_dynamics}

We first investigate the relations between two-body and three-body dynamical systems in the case of linear interaction functions.
Linear dynamics is obviously of critical importance as it often serves as a first approximation and determines the linear stability of critical points for non-linear systems. 
With two-body dynamical systems, the interaction function is given by $f_{ij}(x_i,x_j)=c(x_j-x_i)$ where $c\in\mathbb{R}$ is a scaling constant, and the 
resulting dynamics reads
\begin{align}
\label{eq:lap}
	\dot{x}_i=\sum_{j}A_{ij}c(x_j-x_i)=- c \sum_{j} L_{ij}x_j,
\end{align}
where $L_{ij}=D_{ij}-A_{ij}$ is the network Laplacian.
Here the degree matrix $D_{ij}=\delta_{ij}k_i$ is a diagonal matrix of the degrees $k_i=\sum_{j}A_{ij}$. Eq. (\ref{eq:lap}) naturally arises when modelling continuous-time random walks on networks \cite{masuda_random_2017}, but also for opinion-formation and decentralized consensus, as in the continuous-time DeGroot model \cite{olfati2007consensus}. For undirected, connected networks, the dynamics asymptotically converges to consensus, with a rate determined by the second dominant eigenvalue of the Laplacian.

With three-body interaction systems, the linear interaction function is a linear combination of the diffusive couplings on the two edges $(i,j)$ and $(i,k)$. 
After accounting for the symmetry in $j$ and $k$, one finds the linear interaction function  $$f_i^{\{jk\}}(x_i,x_j,x_k)=c((x_j-x_i)+(x_k-x_i)),$$ where $c$ is again a constant. The three-body dynamical system simplifies enormously in this case, as
\begin{align}
	\dot{x}_i & =  \sum_{jk}A_{ijk}c((x_j-x_i)+(x_k-x_i)) \\
	          & = 2c\sum_{jk}A_{ijk}(x_j-x_i)    
	           =-2c\sum_{j} (L_T)_{ij}x_j, 	 \nonumber
\end{align}
where 
\begin{align}
	(L_T)_{ij}=(D_T - W_T)_{ij}
\end{align}
is the triangle motif Laplacian, defined as the standard Laplacian of a network whose adjacency matrix is
\begin{align}
\label{eqn:rescale}
(W_T)_{ij}=\sum_{k} A_{ijk}.
\end{align}
This rescaled network is thus obtained by weighting  each  edge by the number of triangles to which it belongs. 
The diagonal degree matrix 
\begin{align}
	(D_T)_{ii}=\sum_{kj} A_{ijk}.
\end{align}
counts the number of  triangles around a node.
In other words, a three-body dynamical system can be rewritten as a two-body dynamical system in the case of linear dynamics, after a proper rescaling of the adjacency matrix.
This observation reveals that a genuine multi-body dynamics on hypergraphs requires a non-linear interaction function, and that multi-body interactions are thus not sufficient to produce dynamics that can not be reduced to two-body dynamical systems. 
It is therefore essential to consider the interplay between the interaction function and the multi-body topology of the system.

Before exploring further the dynamics, let us note the connection between the operation \eqref{eqn:rescale} and the `motif matrix' used to uncover communities in higher-order networks \cite{benson_higher_order_2016}. 
Motifs are an important object of study in network science  \cite{milo_network_2002} and are defined as follows.
A \textit{motif} on $k$ nodes is defined by a tuple $(B, P)$, where $B$ is a $k\times k$ binary matrix and $P \subset \{1, 2, .\dots , k\}$ is a set of anchor nodes.
The matrix $B$ encodes the edge pattern between the $k$ nodes.
In \cite{benson_higher_order_2016}, the authors searched to define a generalisation of conductance and of the cut, where the basic unit of interaction would be a specific motif and not an edge.
Essentially, their analysis led to the analysis of a \emph{motif adjacency matrix} 
\begin{align}
	(W_M)_{ij} = \text{number of instances of motifs in }M \\ \text{ containing } i \text{ and }j, \nonumber 
\end{align}
from which a motif Laplacian could be defined. 
The results from \eqref{eqn:rescale} provide a dynamical interpretation of this quantity for triangles, which can readily be generalised to arbitrary $k$-cliques, and could also open the way to the use of random-walk based community detection techniques, such as the Map equation \cite{rosvall2009map} and Markov stability \cite{lambiotte2014random}, for higher-order networks.

\section{Higher-Order Effects of Non-Linear Dynamics}
\label{sec:higher_order_effects}

\subsection{Modelling Consensus Dynamics with Group reinforcement}

In this section, we  introduce and study a specific form of non-linear interaction function, aiming at
modelling consensus dynamics with group reinforcement on hypergraphs. Note that other choices of non-linear interaction functions, akin to Watts threshold models \cite{watts2002simple}, have been considered recently for information spreading \cite{arruda2019social}.
As previously mentioned, multi-body group effects that cannot be reduced to pairwise interactions can appear in various contexts.
In the area of sociology, reinforcing group effects such as peer pressure are a long-standing area of study, for instance in social psychology \cite{asch_effects_1951}.
In the case of binary opinions, for instance, these phenomena emerge when the influence of one opinion on the other in a group depends non-linearly on its popularity. 
Another important mechanism for opinion dynamics is based on homophily \cite{mcpherson2001birds}, and the fact that only sufficiently similar nodes tend to interact with each other \cite{deffuant2000mixing}.
It is important to develop models that capture these  multi-body mechanisms to better understand phenomena such as hate communities \cite{johnson_hidden_2019}, echo chambers and polarisation \cite{bail_exposure_2018} in society.

Motivated by these observations, and inspired by Eq.~\eqref{eqn:pairwise_interaction}, we introduce a three-body consensus model (3CM), with a non-linear interaction function of the form 
\begin{align}
	\label{eqn:our_function}
	f^{\{j,k\}}_{i}(x_i,x_j,x_k)=s(\left|x_j-x_k\right|)\left[(x_j-x_i)+(x_k-x_i)\right],
\end{align}
where we assume the function on each triangle is the same, for the sake of simplicity, and 
with a resulting dynamics for each node $i$ given by
\begin{align}
	\dot{x}_{i}=\sum_{j , k=1}^N A_{ijk} \,s(\left|x_j-x_k\right|) \,\left[(x_j-x_i)+(x_k-x_i)\right].
\end{align}
This expression models, for each triangle $\{i,j,k\}$, the multi-body influence of nodes $j$ and $k$ on node $i$ by the standard linear term $\left[(x_j-x_i)+(x_k-x_i)\right]$ 
modulated by an influence function $s(\left|x_j-x_k\right|)$ of their state differences.
If $s(x)$ is monotonically decreasing,  $j$ and $k$ reinforce their influence on $i$ if they have similar states and hinder each other if they are very different.
This property is reminiscent of non-linear voter models in the case of discrete dynamics \cite{lambiotte_dynamics_2008,molofsky_local_1999, mellor2017heterogeneous}, where the voters change opinion with a probability $p_f$ that depends non-linearly on the fraction $f$ of disagreeing neighbours. 
It is also akin to the Sznajd model modelling the proverb that `United we stand, divided we fall' in sociophysics \cite{stauffer2002sociophysics}. 
In Eq. (\ref{eqn:our_function}), the interaction function is non-linear for non-constant $s(x)$ and captures multi-body effects, as the interactions on a triangle can no longer be split into pairwise interaction functions.
If $s(x)$ is constant, we recover the linear case discussed in \Cref{sec:linear_dynamics}.
Note that our choice of interaction function is clearly symmetric in $i$ and $j$ and therefore congruent with the underlying undirected triangle topology. 
As rotations are isometries  preserving the norm, $s(|x_j-x_k|)(x_j-x_i)$ is rotational and translational invariant. 
Therefore, this property is also true for $f^{\{j,k\}}_{i}(x_i,x_j,x_k)$, which is a sum of two functions of this form.

\begin{figure}
	\centering
	\def\svgwidth{0.5\columnwidth}
	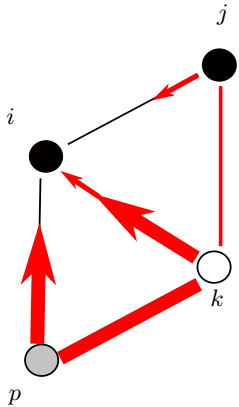
	\caption{The influence on node $i$ due to the interactions on triangles $\{i,j,k\}$ and $\{i,k,p\}$. The value $x_i \in [0,1]$ of the nodes is visualised by a colour gradient between white and black. We consider a monotonically decreasing influence function (e.g. $s(x)=\exp(\lambda x)$ for $\lambda<0$). As a result, nodes $p$ and $k$ reinforce each other as they have similar values, and hence a large $s(\left|x_p-x_k\right|)$. Nodes $k$ and $j$ have instead  distinct values, which leads to a smaller scaling function $s(\left|x_j-x_k\right|)$. As the edge $\{i,k\}$ is part of both triangles, the diffusion on this edge is scaled by both processes.}
	\label{fig:influence_strongandweak}
\end{figure}

A natural choice for the function $s(x)$ is
\begin{align}
	\label{eqn:exponential}
	s(\left|x_j-x_k\right|)=\exp(\lambda |x_j-x_k|),
\end{align}
where the sign of the parameter $\lambda$ determines if the function monotonically decreases or increases and, therefore, if similar or disparate values of $j$ and $k$ accelerate the dynamics on node $i$.
When $\lambda=0$, we recover linear interactions with a constant $s(x)=1$.
\Cref{fig:influence_strongandweak} shows the influences on node $i$ for $\lambda < 0$, i.e. where similar node states reinforce each other.

Another option is the Heaviside function, given by
\begin{align}
	s(|x_j-x_k|)=H(|x_j-x_k|-\phi)=\begin{cases}0 & \mbox{if } |x_j-x_k|<\phi \\
		1 & \mbox{otherwise},
	\end{cases}
\end{align}
which switches between a zero interaction and linear diffusion when the difference of the neighbouring triangle nodes becomes smaller than a threshold $\phi \in (0,1)$.
Note that this function is not positive-definite, so that the dynamics do not necessarily converge to consensus, with all nodes have the same value, asymptotically.

We will examine the dynamics governed by these two different forms of interaction functions in more detail in our numerical experiments.
Until then, we will assume that $s(x)$ is an arbitrary scalar function.

\subsection{Derivation of a Weighted, Time-dependent Laplacian}
\label{sec:higher_order_effects:weighted_laplacian}

In \Cref{sec:linear_dynamics}, we showed that, in the case of linear interactions, a three-body dynamical system can be rewritten as a  standard two-body dynamical system, defined on a network where the weight of an edge is the number of triangles to which it belongs. Let us explore how this result extends for 3CM.
Recall that we assumed that the adjacency tensor $A_{ijk}$ is symmetric.
We define $\mathcal{I}_{ij}$ as the index-set of nodes that form a triangle with nodes $i,j$.
Note that $\mathcal{I}_{ij}=\emptyset$ if no triangle exists or, equivalently, if the edge $\{i,j\}$ does not exist.
We now define the weighted adjacency matrix $\mathfrak{W}$ as
\begin{align}
\label{eq:hello}
	(\mathfrak{W})_{ij}= \sum_{k}A_{ijk}s(\left|x_j-x_k\right|) =\sum_{k \in \mathcal{I}_{ij}}s(\left|x_j-x_k\right|).
\end{align}
The entries $(\mathfrak{W})_{ij}$ measure the three-body interactions on node $i$ over edge $\{i,j\}$. The matrix verifies  $\text{diag}(\mathfrak{W})=0$, because $\mathcal{I}_{ii}=\emptyset$ as we excluded triangles formed by self-loops.
The corresponding degree matrix measuring the total three-body influence on node $i$  is defined as
\begin{align}
	(\mathcal{D})_{ii}=  \sum_{jk}A_{ijk}s(\left|x_j-x_k\right|) = \sum_{j} \mathfrak{W}_{ij}
\end{align}
and the corresponding  Laplacian is then given by
\begin{align}
	\label{eqn:weighted_laplacian}
	(\mathcal{L})_{ij}= (\mathcal{D}-\mathfrak{W})_{ij}.
\end{align}
Using \eqref{eqn:weighted_laplacian}, we can rewrite the dynamics as
\begin{align}
\label{eq:ren}
	\dot{x}_i & = \sum_{jk}A_{ijk} \,s(\left|x_j-x_k\right|)\,((x_j-x_i)+(x_k-x_i)) \nonumber \\
	          & = 2\sum_{jk}A_{ijk}\, s(\left|x_j-x_k\right|) \, (x_j -x_i)           \nonumber \\
	          & = 2\sum_{j}\mathfrak{W}_{ij}(x_j-x_i)						
	=  -2\sum_{j}\mathcal{L}_{ij}x_j.											
\end{align}
The 3CM can thus also be rewritten in terms of the Laplacian of a network, with the essential difference that this network is now time-dependent, through its dependence of the adjacency matrix $\mathfrak{W}=\mathfrak{W}(t)$ on the dynamical node states $x_i=x_i(t)$ in (\ref{eq:hello}).
We drop this notation for simplicity, simply using $\mathfrak{W}$ from now on, but note that the weighted Laplacian is indeed constant in time in the linear case.
The weighted Laplacian is the matrix representation of the non-linear dynamics and therefore the analogue of the motif Laplacian, introduced in \Cref{sec:linear_dynamics} for linear dynamics.
Even if this reformulation suggests that the dynamical system may be rewritten as a pairwise system, the entries of the weighted Laplacian are dependent on the node states of the neighbouring triangle nodes, reflecting the genuine three-body interactions of 3CM.

\subsection{Conservation of the Average Node State}
\label{sec:higher_order_effects:conservation}

Our objective is to determine if, and how, 3CM asymptotically reaches consensus. The identification of conserved quantities is usually an essential step to understand the properties of dynamical systems.
In the case of  consensus on networks, it is well-known that the average state at time $t$, 
\begin{align}
	\bar{x}(t)=\frac{1}{N}\sum_{i=1}^Nx_i(t),
\end{align}
is conserved under general conditions. Consider a two-body dynamical system described by
\begin{align}
	\dot{x}_i(t) &= \sum_{j=1}^N A_{ij}f_{ij}(x_i(t),x_j(t)) \\ &= \sum_{j=1}^N A_{ij}g(x_j(t)-x_i(t)). \nonumber
\end{align}
The initial average $\bar{x}(0)$ is conserved if the derivative $\dot{\bar{x}}(t) =\frac{1}{N}\sum_{i,j=1}^N A_{ij}g(x_j(t)-x_i(t))$ is zero for all times.
This is true if the adjacency matrix $A_{ij}$ of the underlying graph is symmetric and the interaction function $g(x)$ is odd, which is verified for quasi-linear dynamics.

Consider now a three-body dynamical system with adjacency tensor $A_{ijk}$ and the interaction function $f(x)$
\begin{align}
	\dot{x_i}(t) & = \sum_{jk}A_{ijk}f(x_i(t),x_j(t),x_k(t)) \\
	&= \sum_{jk}A_{ijk}g((x_j(t)-x_i(t))+(x_k(t)-x_i(t))). \nonumber
\end{align}
where the form of the interaction function  $g$ ensures that the dynamics are rooted in $i$ and symmetric in $j$ and $k$, in congruence with our model.
Let $\Pi(i,j,k)$ be the set of all permutations of the three indices.
Using this notation, we can conclude that the derivative $\dot{\bar{x}}(t)= \frac{1}{N}\sum_{i,j,k=1}^N A_{ijk}g((x_j(t)-x_i(t))+(x_k(t)-x_i(t))) $ is zero for all times if $A_{\pi}=A_{\tau}$ for all permutations $\pi,\tau \in \Pi(i,j,k)$ and  $g(x_j(t)+x_k(t)-2x_i(t))+g(x_i(t)+x_k(t)-2x_j(t)) + g(x_j(t)+x_i(t)-2x_k(t)) = 0$.
This is the case for an undirected three-body interactions  (such that $A$ is symmetric in all dimensions) if we choose $g(x)$ to be a linear function.
We thus conclude that in three-body dynamical systems, linear dynamics conserves the average state of the system.

For non-linear dynamics, the conservation of the average state is in general not verified. For 3CM, where the interaction function takes the specific form $f(x_i,x_j,x_k)=s(\left|x_j-x_k\right|)((x_j-x_i)+(x_k-x_i))$, the change in the average state can be written as
\begin{align}
	\dot{\bar{x}}(t)= \frac{1}{N}\sum_{i=1}^N\dot{x_i}(t) & = \frac{1}{N}\sum_{i,j=1}^N  2\sum_{j}\mathfrak{W}_{ij}(x_j(t)-x_i(t)).
\end{align}
Trivially, the average state is conserved if $\mathfrak{W}$ is symmetric.
For all $i,j$,
\begin{align}
	\mathfrak{W}_{ij}                                          & =\mathfrak{W}_{ji}                                                      \\
	\Leftrightarrow
	\sum_{k \in \mathcal{I}_{ij}}s(\left|x_j(t)-x_k(t)\right|) & = \sum_{k \in \mathcal{I}_{ij}}s(\left|x_i(t)-x_k(t)\right|), \nonumber
	\label{eqn:symmetry}
\end{align}
which is only  true, for all times, if the influence function is constant, $s(x)=c$, that is when the dynamics is linear.
The weighted matrix $\mathfrak{W}_{ij}=c(W_T)_{ij}$ is then  the motif adjacency matrix scaled by the constant $c$, in agreement with the results in \Cref{sec:linear_dynamics}.
Other symmetries are visible in the weighted matrix: if every entry of the $i$-th row of $\mathfrak{W}_{ij}$ is the same, node $i$ has the same incoming multi-body effect from all its neighbours.
If the same holds for the $i$-th column, node $i$ has equal outgoing multi-body effect over all triangle edges.

When the dynamics is non-linear and $s(x)$ is not a constant,  $\mathfrak{W}(t)$ and the weighted Laplacian are  time-dependent. If, at a certain time $t$,  $\mathfrak{W}(t)$ is symmetric, the derivative of the average state vanishes locally in time. 
This happens whenever the multi-body effects on the triangles are in balance with each other, as in Eq.~\eqref{eqn:symmetry}.
This balance is determined by how many triangles a node is a part of, what states the nodes in these triangles are in, and the form of the influence function $s(x)$.
Let us assume that the system is initially such that $\mathfrak{W}(0)$ is symmetric, and  examine how the symmetry is affected for later times $t>0$.
As we are working with undirected and thus symmetric three-body interactions, the dynamics are symmetric on each triangle.
The triangle topology and the interaction function itself play a role in the evolving process, as they are encoded in the weighted matrix.
Therefore, we need a three-body topology that is symmetric concerning the form of $s(x)$ in addition to a symmetric initial adjacency matrix for this property to be conserved.
Otherwise, the average state shifts according to the interplay of influence function $s(x)$, initialisation, and network topology.

\subsection{Influence of Initialisation in the Mean-Field}
\label{sec:higher_order_effects:meanfield}

In order to investigate how the non-linearity of 3CM affects the conservation of the average state, we first consider a mean-field approach. 
We consider a fully-connected hypergraph $\mathcal{H}$ with $N$ nodes $V(\mathcal{G})=\{1, \dots, N\}$, such that each triplet of distinct nodes is connected by a triangle.
The generalised adjacency tensor is thus
\begin{align}
	A_{ijk}=\begin{cases}
		1 & \text{ if } i\neq j \neq k \\
		0 & \text{ otherwise.}
	\end{cases}
\end{align}
We have that $\mathcal{I}_{ij}=V(\mathcal{G})/\{{i,j\}}$ with  $|\mathcal{I}_{ij}|=(N-2)$ for all $i,j$ with $i \neq j$ and $\mathcal{I}_{ij}=\emptyset$ for $i=j$.
The dynamics at time $t$ are then given by
\begin{align}
	\dot{x_i}(t) & = 2\sum_{j}\mathfrak{W}_{ij}(t)(x_j(t)-x_i(t)) \nonumber \\ 
	&=2\sum_{j=1}^N(x_j(t)-x_i(t))\sum_{k \in \mathcal{I}_{ij}}s(\left|x_j(t)-x_k(t)\right|).
\end{align}

The average state of those dynamics is represented by $\bar{x}(t)=\frac{1}{N}\sum_{i=1}^Nx_i(t)$ and is conserved if $\mathfrak{W}(t)$ is symmetric.
In the fully connected case this condition holds if 
\begin{align}
	\label{eqn:equality}
	\sum_{k \in \mathcal{I}_{ij}}s(\left|x_j(t)-x_k(t)\right|)  = \sum_{k \in \mathcal{I}_{ij}}s(\left|x_i(t)-x_k(t)\right|)
\end{align}
for all nodes $i$ and $j$ in the network.
This means that the multi-body effects of the interactions must be balanced for all nodes in a fully connected system.
This condition is the same as in the general case, see Eq.~\eqref{eqn:symmetry}, although the mean-field ignores the additional effect of topological distributions of the triangles on the graph.
The equality only holds in general for linear dynamics if $s(x)$ is constant. Otherwise, a shift of the initial state may be observed, depending on the initialisation.
As an illustration,  consider a situation where the number of nodes is even and when the initial values on the nodes is binary, that is either zero or one.  Condition (\ref{eqn:equality}) will be satisfied only if the initial configuration is balanced, that is when  $\bar{x}(0)=0.5$, in which case that quantity is conserved in time. 
In contrast, if the initial configuration is unbalanced, there will necessarily exist edges over which (\ref{eqn:equality}) does not hold, and the average state is expected to evolve in time. 
 If $s(\left|x_j-x_k\right|) $ is given by a decreasing function, that is when similar nodes reinforce each other, the deviation from  $0.5$ is expected to grow in time, with a drift towards the majority.
In contrast, if the influence function is such that dissimilar nodes reinforce each other, one expects to observe a drift to the balanced state $0.5$.

\begin{figure*}
	\centering
	\begin{subfigure}{0.32\textwidth}
		\caption{$\lambda=-1$}
		\includegraphics[width=\textwidth]{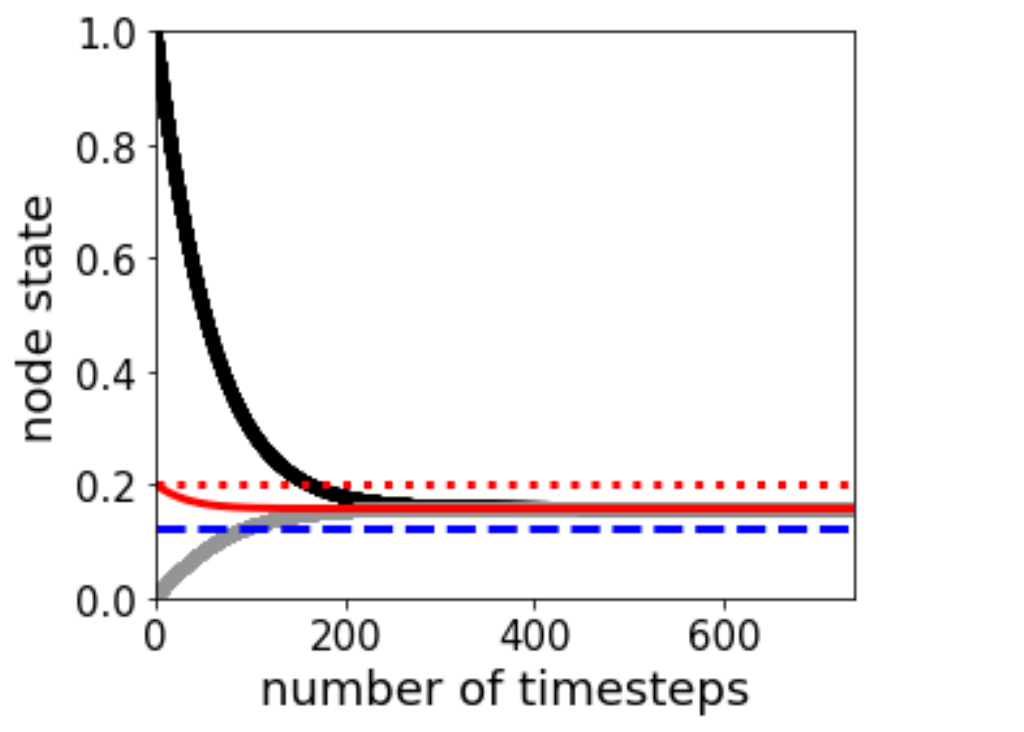}
	\end{subfigure}
	\begin{subfigure}{0.32\textwidth}
		\caption{$\lambda=0$}
		\includegraphics[width=\textwidth]{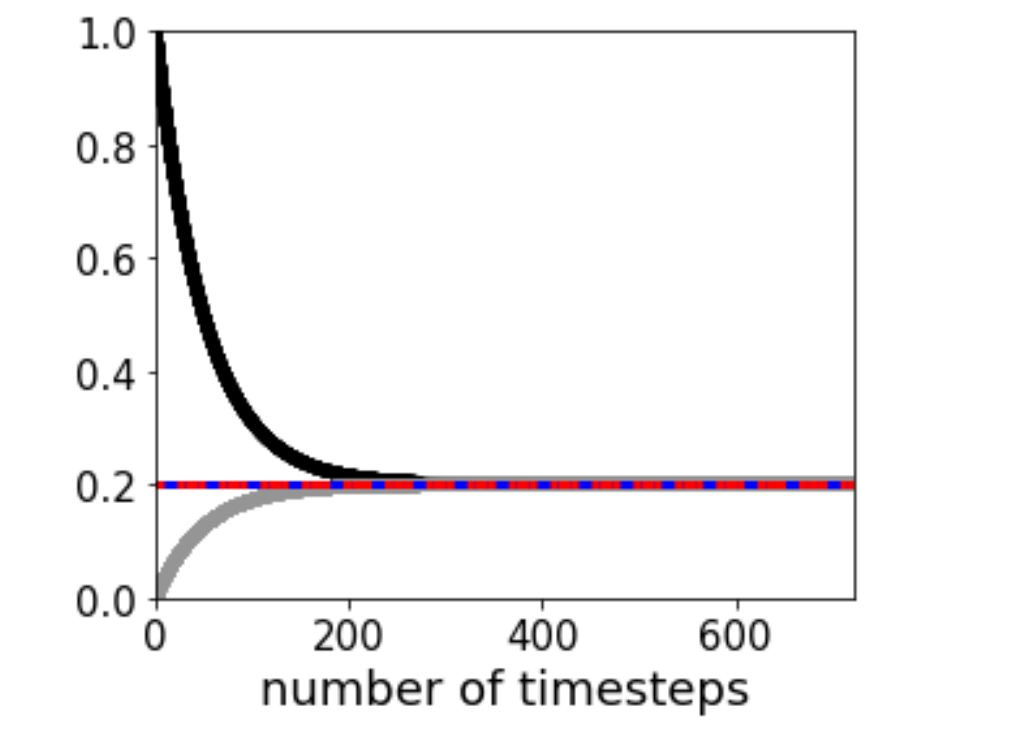}
	\end{subfigure}
	\begin{subfigure}{0.32\textwidth}
		\caption{$\lambda=1$}
		\includegraphics[width=\textwidth]{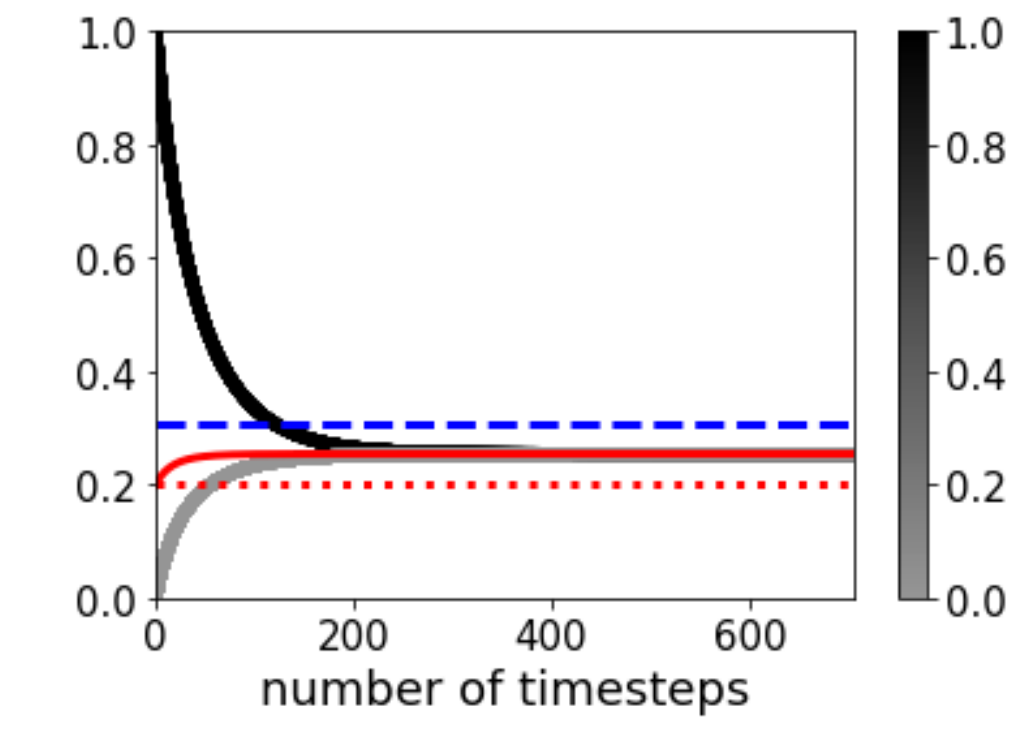}
	\end{subfigure}
	\caption{
		An asymmetric initialisation, with $\bar{x}(0)=0.2$, may shift the average node state in 3CM for fully-connected hypergraphs. The interaction function is $s(x)=\exp(\lambda x)$. 
		For $\lambda<0$ in (a), the dynamics exhibits a drift towards the majority as similar node states reinforce each other. The opposite effect occurs for $\lambda>0$ in (c), as the dynamics exhibits a drift towards balance.
		The average state is conserved for $\lambda=0$ in (b), as expected for  linear dynamics. 
		Dotted red lines indicate the initial value of the  average node state. Black (grey) solid lines represent the evolution of the state of nodes whose initial configuration is one (or zero). Dashed blue lines are the final state approximation, $\bar{x}^p$.}
	\label{fig:simulations_meanfield_biased}
\end{figure*}

To validate these findings,
we have performed numerical simulations of 3CM on a fully connected hypergraph of 100 nodes, and used the exponential influence function $s(x)=\exp(\lambda x)$ with $\lambda=-1$ for a decreasing function, $\lambda=1$ for a growing function and $\lambda=0$ for a constant.
Considering different initial distributions of the node states, we have compared:
\begin{enumerate}
	\item a random initialisation with uniform distribution $\mathcal{U}([0,1])$ ($\bar{x}(0)=0.5$)
	\item a symmetric, binary initialisation where $50\%$ of the nodes are initialised with value $0$ and $50\%$ with value $1$ ($\bar{x}(0)=0.5$)
	\item an asymmetric, binary initialisation where $80\%$ of the nodes are initialised with value $0$ and $20\%$ with value $1$ ($\bar{x}(0)=0.2$)
\end{enumerate}
In the first two cases we do not observe any shift in the average state, as expected.
However, we do see a shift for the asymmetric initialisation, as shown in \Cref{fig:simulations_meanfield_biased}.
The simulations confirm that the average state is conserved for linear dynamics ($\lambda=0$) and multi-body effects only occur for non-linear interaction functions with $\lambda \neq 0$.
For $\lambda < 0$ we observe a shift towards the majority, resulting in an asymptotic average smaller than the initial value of $0.2$ and growing imbalance.
For $\lambda < 0$ we see the opposite phenomenon, with a shift of the average opinion towards balance.

Finally, let us propose a simple method to approximate the asymptotic value of the average state based on the initial configuration. To do so, we estimate the dynamical importance $w_i$ of a node $i$ based on the initial configuration as
\begin{align}
	w_i(t) & =\frac{\text{influence of node } i}{\text{total weight in the system}}                            \nonumber         \\
	       & =\frac{\sum_{j,k=1}^NA_{ijk}s(\left|x_j(t)-x_i(t)\right|)}{\sum_{i,j,k=1}^N A_{ijk} s(\left|x_j(t)-x_i(t)\right|)}.
\end{align}
The asymptotic value of $\bar{x}$ is then obtained by one explicit Euler step of the dynamics from the initial configuration  $\bar{x}(0)$ 
\begin{align}
	\label{eqn:predicted_mean}
	\bar{x}^p=\bar{x}(0)+ \sum_{i=1}^Nw_j(0)(x_j(0)-x_i(0)).
\end{align}
The simulations in \Cref{fig:simulations_meanfield_biased}  also display the predicted value \eqref{eqn:predicted_mean}, which  correctly identifies the direction of the shift.

\subsection{Influence of the Triangle Network Topology}
\label{sec:higher_order_effects:influence_topology}

As a next step, let us go beyond the mean-field and consider non-trivial topologies.
A critical aspect is the presence of asymmetries in the topology, in particular of two connected clusters.
Consider two fully connected clusters that consist of the same number of nodes, and the clusters are connected by a set of triangles.
Since triangles are made of three nodes and three edges, each triangle defines an asymmetric connection between the clusters.
Two edges connect nodes of different clusters (\textit{outer-edges}) and one edge connects two nodes within one cluster (\textit{inner-edge}).
We call a connecting triangle between cluster $A$ and cluster $B$ `directed towards cluster $B$' if $A$ contains the inner-edge.

\begin{figure}
	\centering
	\includegraphics[width=0.45\textwidth]{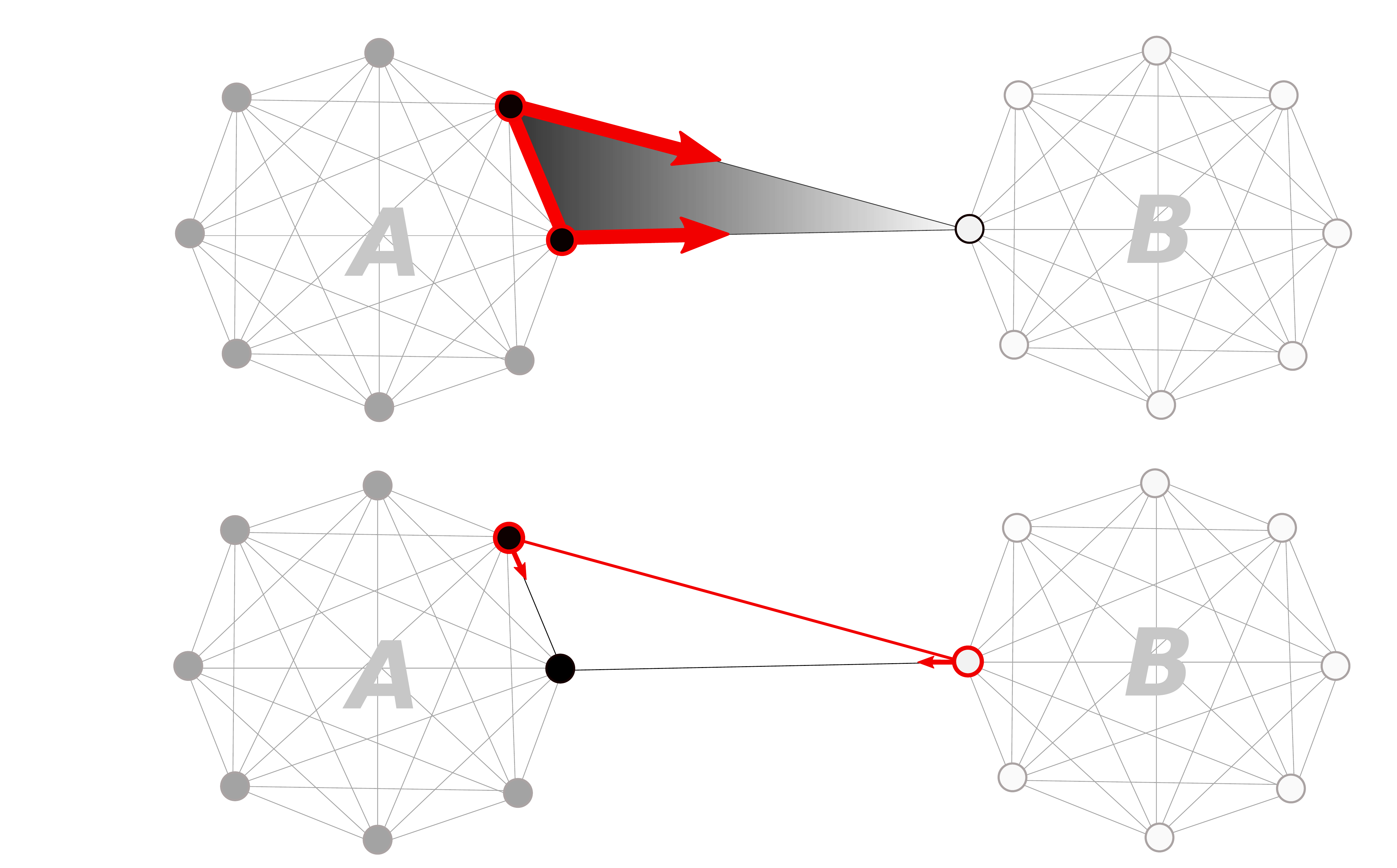}
	\caption{
		If we consider a binary initialisation of the two clusters, here in black and white, and a  triangle directed towards cluster $B$ (top), the consensus in cluster $A$, and thus on the inner-edge, accelerates the rate of change of the neighbour in $B$. 
		In contrast, the node-state difference between the clusters, and thus on the outer-edges, is maximal, which slows down the effect of cluster $B$ on $A$.
		}
	\label{fig:binary_cluster}
\end{figure}

The effect of this topological construction becomes clear if we additionally take the initial node states into account.
Consider the case of a binary initialisation, shown in \Cref{fig:binary_cluster}, with the nodes in cluster A in the initial state  $x_A(0)=0$ and the nodes in cluster B in the initial  state $x_B(0)=1$. We consider a positive-definite, decreasing  influence function $s(x)$, so that similar states reinforce each other. Moreover, the configuration is such that there is one single triangle between $A$ and $B$, and it is directed towards cluster $B$.
Due to the consensus in cluster A and the fact that it contains an inner-edge, the diffusion on the two outer-edges of the connecting triangle is accelerated.
On the contrary, the influence is inhibited in the opposite direction, as the two outer-edges damp the diffusion because of the large state difference between the clusters. For this reason, one expects the initial value in $A$ to dominate that in $B$ and thus to determine the asymptotic consensus value. 
Despite the lack of direction of edges, the process leads to an asymmetric flow of influence from one cluster to the other \footnote{Note that the direction of the dynamics reverses if we consider an increasing influence function such that dissimilar node states reinforce each other.}. 

\begin{figure}
	\centering
	\begin{subfigure}{0.49\linewidth}
		\includegraphics[width=\linewidth]{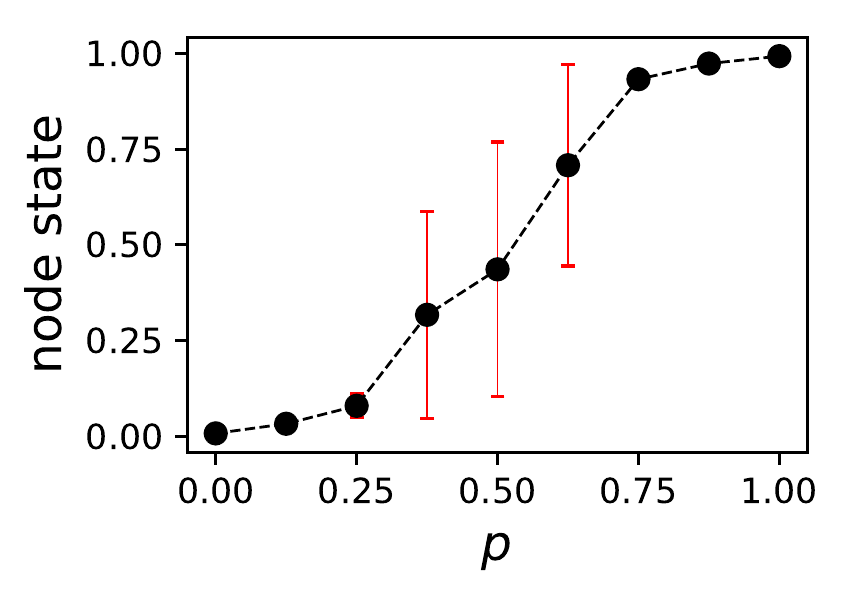}
		\caption{Consensus value}
	\end{subfigure}
	\begin{subfigure}{0.49\linewidth}
		\includegraphics[width=\linewidth]{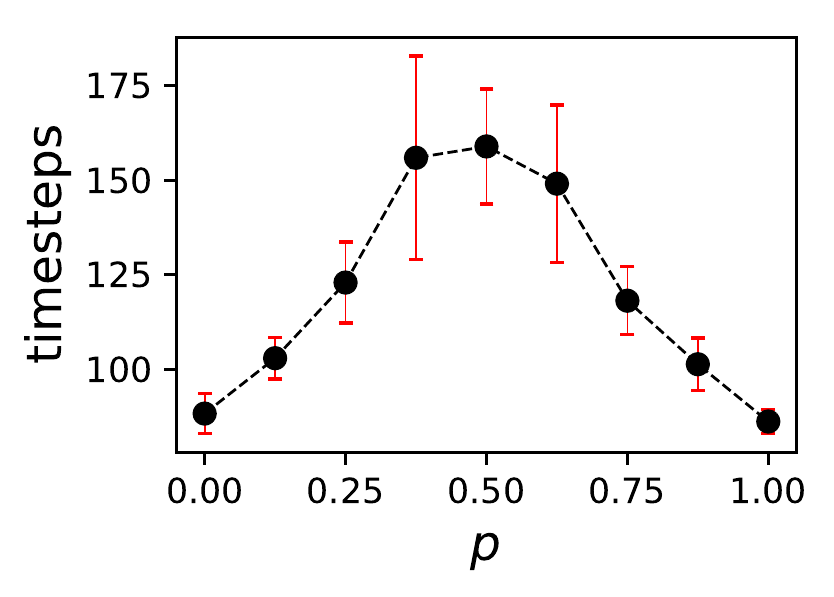}
		\caption{Timesteps until consensus}
	\end{subfigure}
	\caption{Simulations of 3CM on two inter-connected clusters of 10 nodes, with  the interaction function  $s(x)=\exp(-100x)$ (see main text for a complete description).
		(a) The final consensus value, averaged over 20 simulations, where the error bars display one standard deviation.
		As the fraction of triangles directed from cluster A to cluster B increases, so does the consensus value towards the initial state in cluster A.
		(b) The rate of convergence is significantly faster when the initial configuration is very asymmetric, that is extreme values of $p$.
		}
	\label{fig:p_exp}
\end{figure}

In order to analyse this mechanism more quantitatively, we perform numerical simulations on two fully connected clusters,  each consisting of $10$ nodes, with the binary initialisation  specified above.
We then connect the clusters with $80$ randomly placed triangles, such that a fraction  $p \in [0,1]$ of triangles are directed towards cluster A and the rest towards cluster B.
We first  examine the influence of the directedness parameter $p$.
For that purpose, we take the interaction function $s(x)=\exp(\lambda x)$ with $\lambda =-100$, so that   pairs of similar nodes exert an overwhelmingly strong influence on other nodes.
We show the results of averages over 20 simulations in \Cref{fig:p_exp}.
In (a), we observe a shift in the final consensus value towards the initial value in cluster $A$ or in cluster $B$, depending on the percentage of triangles directed away from that cluster.
This is expected for the above argument, as the connection is maximally directed towards cluster $B$ for $p=0$ and towards $A$ for $p=1$.
The asymmetry also influences the rate of convergence towards consensus, as shown in (b).
More asymmetric configurations lead to a faster rate of convergence.
The simulations also reveal higher fluctuations in the asymptotic state for values close to $p=0.5$.
This result indicates that the process is  sensitive to even small deviations from balance in the initial topology, which can lead to large differences in the consensus value.

\begin{figure}
	\centering
	\begin{subfigure}{0.8\linewidth}
		\includegraphics[width=\linewidth]{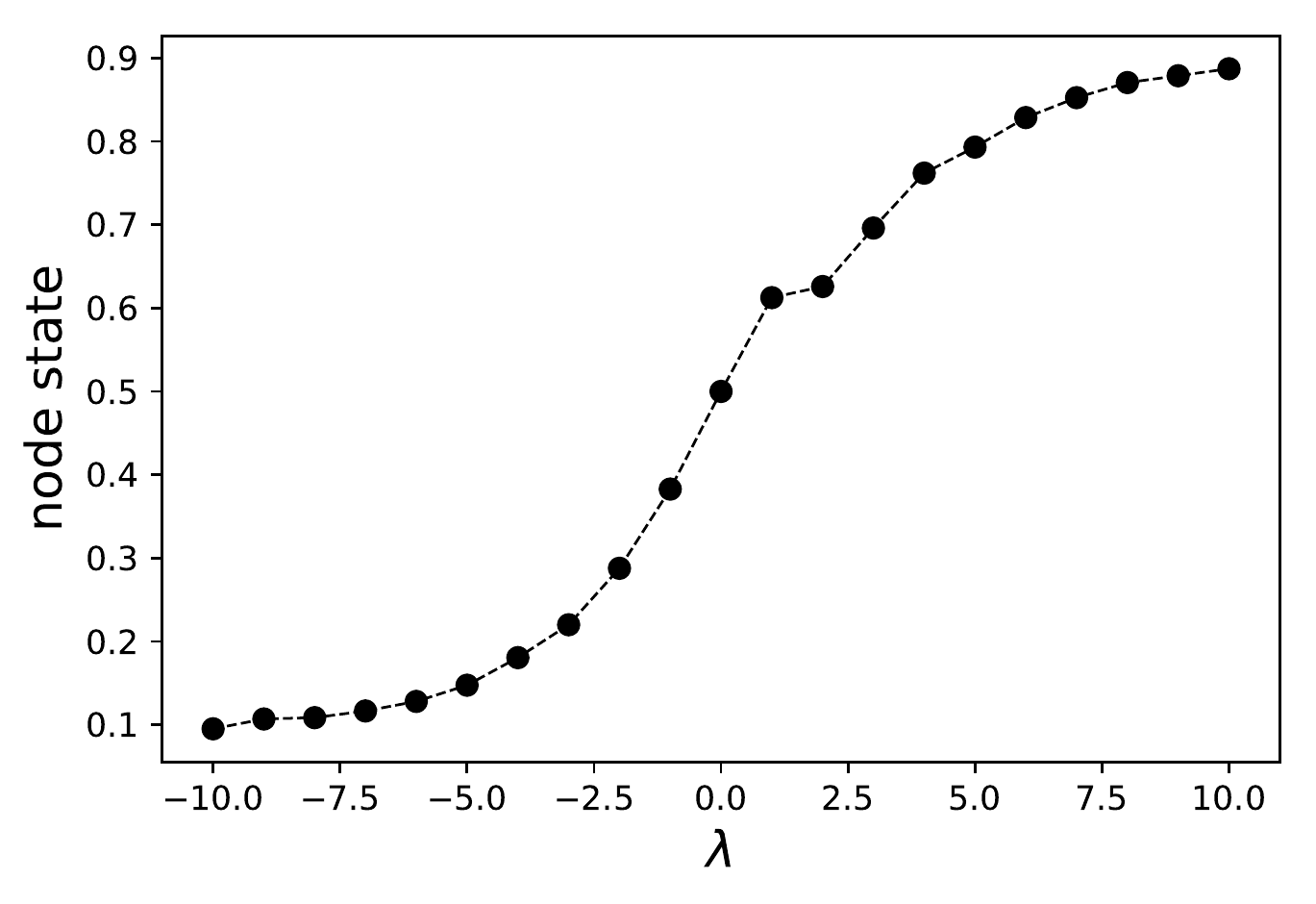}
	\end{subfigure}
	\caption{
	The final consensus value of the two cluster system (with $p=0$), dependent on the parameter $\lambda$.
	As the connection is fully directed towards cluster B, it depends on $\lambda$ if the nodes in cluster A are reinforcing each others' influence which leads to directed dynamics towards cluster B ($\lambda<0$) or inhibit each other ($\lambda>0$) which leads to the contrary effect.
	Therefore, the consensus value shifts towards the mean of cluster B with growing $\lambda$.
	For $\lambda=0$ we have linear dynamics and the initial average $0.5$ is conserved.
	}
	\label{lambda_exp}
\end{figure}

In a second set of experiments, we examine  the impact of the choice of interaction function.
We first consider a maximally directed connection towards cluster $B$ ($p=0$), in order to amplify possible differences, and the influence function $s(x)=\exp(\lambda x)$.
In \Cref{lambda_exp}, we plot the consensus value as a function of $\lambda$, showing a transition from the initial value in cluster $A$ to that in cluster $B$, as expected.
For $\lambda <0$ ($\lambda >0$), the consensus is biased towards the initial value in $A$  ($B$).
For $\lambda = 0$, the dynamics is linear and the initial average state $\bar{x}(0)=0.5$ is  conserved.

\begin{figure}
	\centering
	\includegraphics[width=\linewidth]{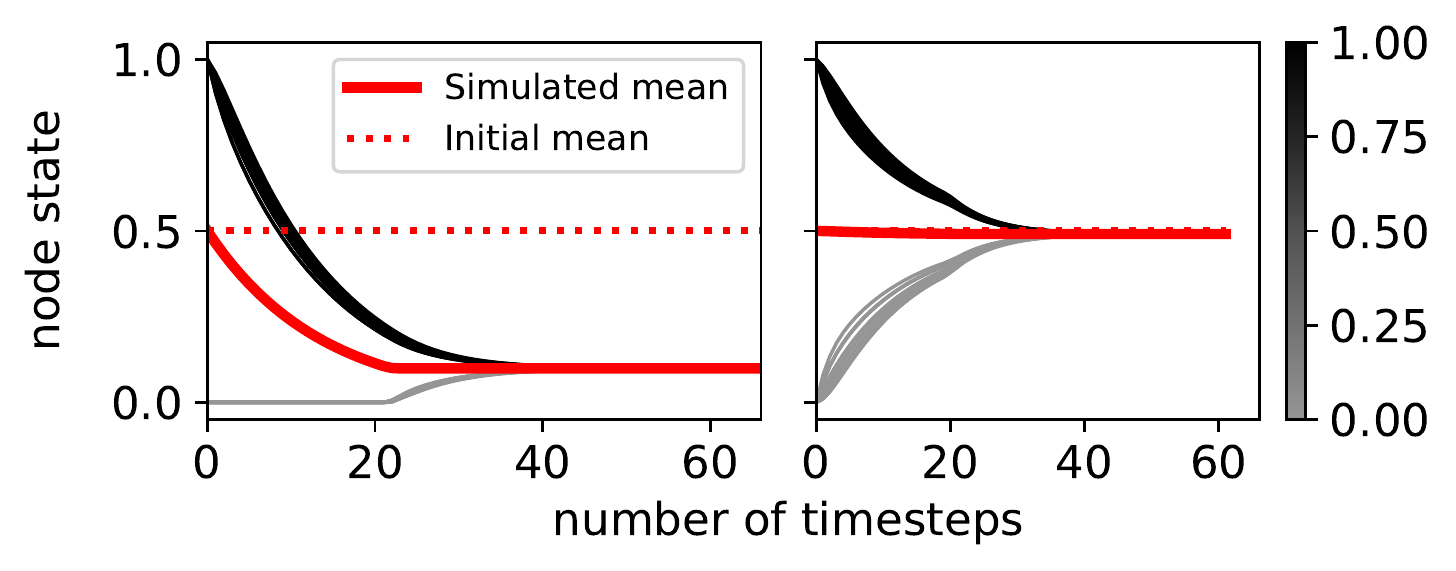}
	\caption{Time evolution of the node states for a Heaviside function with $\phi=0.2$.
		For $p=0.0$ (left), only diffusion from cluster $A$ towards $B$ is enabled, until the threshold of the Heaviside threshold $\phi=0.2$ is reached.
		The dynamics then become linear and the average state becomes conserved.
		For $p=0.5$ (right), the dynamics are initially symmetric, as the directedness of the connection of the two clusters is balanced, and the dynamics are simultaneously switched on.}
	\label{fig:heaviside_exp}
\end{figure}

We also consider a more extreme choice of interaction function,  the Heaviside function, $s(|x_j-x_k|)=\text{H}(|x_j-x_k|-\phi)$ with the threshold $\phi \in (0,1)$. 
This function switches on and off the impact of $f_i^{\{jk\}}(x_i,x_j,x_k)$ on node $i$ depending on the node state difference $\left|x_j-x_k\right|$.
If the interaction  is switched on, it is  linear.
This property is reminiscent of the bounded confidence model \cite{blondel_krauses_2009}, which makes the current framework an extension of it to three-body interactions.
In \Cref{fig:heaviside_exp}, we show the simulation results for $\phi=0.2$.
As the difference between the two clusters is initially larger than $\phi$, the diffusion is only switched on in the direction of the connecting triangle (so towards $B$ for $p=0$).
Therefore, only nodes of cluster $B$ change their value initially as shown in \Cref{fig:heaviside_exp} (a). 
As soon as the distance between the two clusters is smaller than $\phi$, the dynamics becomes linear and the asymmetry of the dynamics disappears.
For $p=0.5$ the dynamics are symmetric as the direction of the triangles is balanced.
Note that there is no shift in the average in this case, as we previously observed for $s(x)=\exp(\lambda x)$. 
Since the dynamics are simultaneously switched on and are then linear, the different realisations of the directing triangles between the clusters do not impact the dynamics.

\begin{figure}
	\centering
		\includegraphics[width=\linewidth]{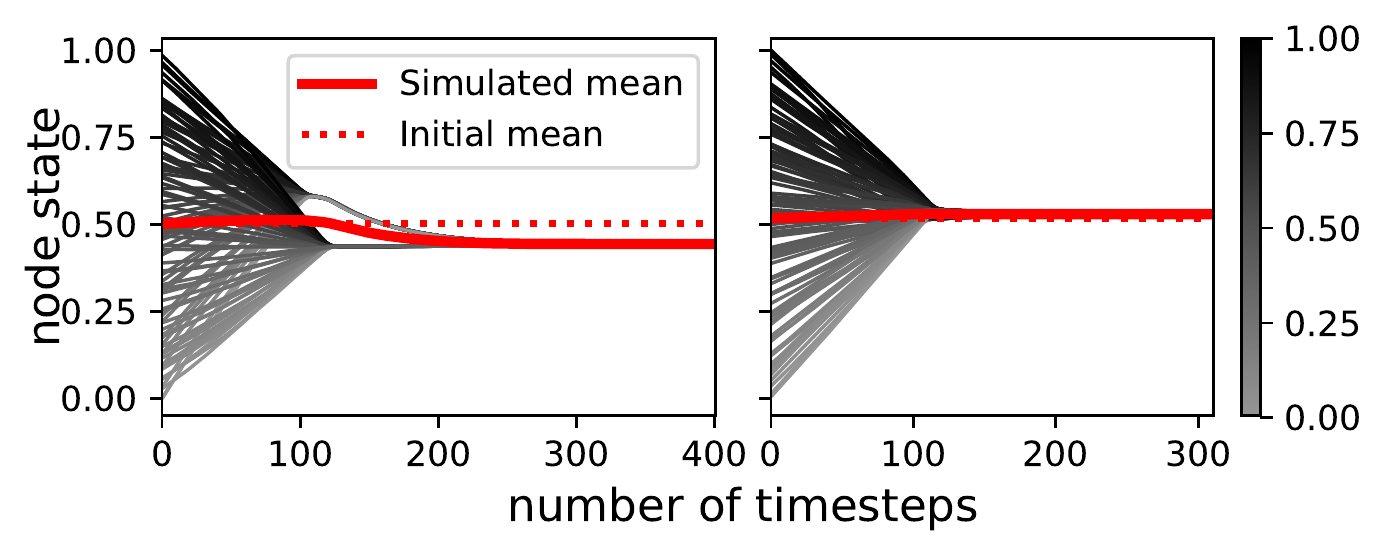}
	\caption{
		We consider two clusters with 50 nodes each and an uniformly random intialisation $\mathcal{U}([0,1])$.
		We connect the two clusters with 400 triangles, with $p=0$ (left), $p=0.5$ (right).
		As before, the value of $p$ quantifies the asymmetry of the topology.
		We observe that the initial imbalance of the node-states in the clusters is reinforced for $p \neq 0.5$, even for very small deviations from an initial average of $\bar{x}(0)=0.5$.
		}
	\label{fig:random_ini}
\end{figure}

In a third experiment, we consider the effect of initialisation on the dynamics. 
As seen in \Cref{sec:higher_order_effects}, conservation of the average node state can be achieved if the multi-body effects are balanced on each node, which depends on  node states and the three-body structure simultaneously.
In the mean-field, we had a fully connected, symmetric topology and therefore an initial state average of $\bar{x}(0)=0.5$ was sufficient.
To achieve a balanced initialisation for the clustered topology, the initialisation has to be chosen to reflect the topological structure, and thus the parameter $p$.
This is a difficult task as the system is sensitive to small deviations from a symmetric topology for certain influence functions, as shown in \Cref{fig:p_exp}.
We would expect to get close to a balance if we initialise the system randomly.
We examine the roles of fluctuations in the initialisation in \Cref{fig:random_ini}, where each node is initially assigned a random number in $[0,1]$, so that 
$\bar{x}(0)=0.5$.
We use the influence function $s(x)=\exp(\lambda x)$ with $\lambda = -100$.
The value of $p$ determines if we have an asymmetric topology ($p=0$) and therefore deviations from the initial mean inside clusters are reinforced (\Cref{fig:random_ini}, left), or if we have almost symmetric topology ($p=0.5$), which leads to a conservation of the mean for small deviations because the system is sufficiently balanced (\Cref{fig:random_ini}, right).
Note that, for $p=0$, the system first reaches a local consensus inside each cluster, before attaining the global consensus asymptotically. 

\begin{figure}
	\centering
	\begin{subfigure}{\linewidth}
		\includegraphics[width=\linewidth]{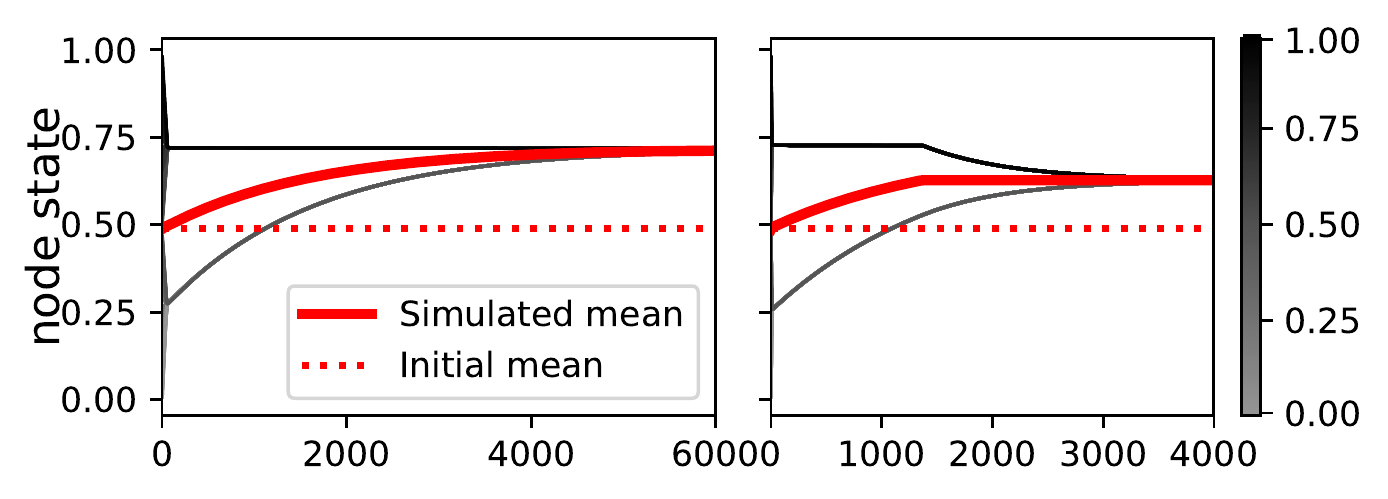}
	\end{subfigure}
	\begin{subfigure}{\linewidth}
		\includegraphics[width=\linewidth]{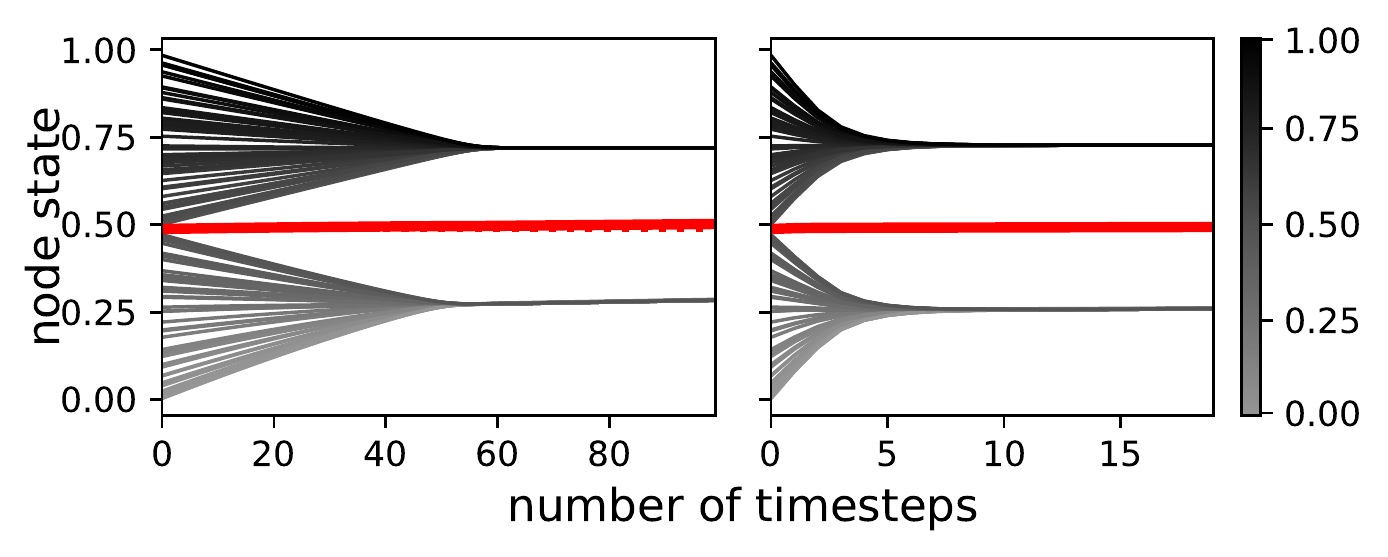}
	\end{subfigure}
	\caption{
		Dynamics of two clusters $A$ and $B$ connected with $p=1$, e.g. with triangles towards nodes in $A$, and initialised with uniform distributions over separate intervals $I_A, I_B$ with $I_A \cap I_B = \emptyset$. The left figures correspond to the exponential interaction function $s(x)=\exp(\lambda x)$ for $\lambda=-100$, and the right to a Heaviside function with threshold  $\phi=0.2$.
		We observe a timescale separation with a fast, symmetric dynamics inside the clusters, followed by a slow, asymmetric dynamics between the clusters. The fast dynamics is shown in the bottom figures, with qualitatively similar results for both interaction functions. The top figures show a shift towards cluster $B$ for the slow dynamics. For the Heaviside function, the process becomes linear when the values in the two clusters are less separated than the Heaviside-threshold.
	}
	\label{fig:timescale_sep}
\end{figure}
As a last set of experiments, we investigate the emergence of different time scales in the dynamics, that is a fast convergence of states inside clusters, followed by a slower convergence towards global consensus. 
To do so, we initialise nodes in different clusters uniformly randomly over separate intervals, such that nodes of cluster $A$ have random values in the interval $I_A=[0,0.5]$ and those of cluster $B$ in $I_B=[0.5,1]$.
In this setting, the initial averages in the two clusters are far apart.
The clusters are internally fully-connected and inter-cluster-dynamics are generally more inhibited due to the difference of the initial cluster means.
As a result, we first observe a fast symmetric dynamics within the clusters towards their means (\Cref{fig:timescale_sep}, bottom) and then a slower dynamics between the two clusters (\Cref{fig:timescale_sep}, top).
However, the outcome of this process critically depends on the interaction function. For $s(x)=\exp(\lambda x)$, with $\lambda=-100$, we observe an asymmetric shift towards cluster B for $p=1$, as shown in \Cref{fig:timescale_sep} (left). 
If we consider the Heaviside function instead, the dynamics between the clusters show a similar asymmetry as in the exponential case until the two cluster means are less separated than the Heaviside threshold $\phi=0.2$. As shown in \Cref{fig:timescale_sep} (right), the dynamics between the clusters then become linear and symmetric.

Together, these results show the importance of the initial distribution of the node states, their alignement with the modular structure of the hypergraph, and also the specific shape of the non-linear interaction function.

\section{Discussion}
\label{sec:discussion}

Network science provides a powerful framework for the modelling and description of interacting systems. Its strength comes from its minimalism and generality, dissecting the notion of connectivity into core elements, nodes and edges, that may then be combined to form indirect connections. Yet, interactions may not always be decomposed into a sum of pairwise edges, and the atomic unit of interaction may thus involve more than two nodes. In this paper, we have explored a simple model, 3CM, for consensus in order to identify the impact of three-body interactions on dynamics.

As a first step, we have clarified the difference between the model of the multi-body structure of a system, and the model of its multi-body dynamics. The distinction is most apparent in the case on linear consensus models, whose dynamics can be reduced to a two-body dynamical system even when they are defined on a higher-body hypergraph. In other words, the interaction needs to be non-linear for genuine, non-reducible multi-body dynamical phenomena to emerge. As a second step, we have introduced a non-linear interaction function inspired by models in opinion dynamics, in which two nodes in a triangle either reinforce or inhibit each other's effect on the third node depending on the own similarity. Note that 3CM is quadratic in the node states and that it may be seen as a first-order correction to linear, two-body dynamics. We have shown that the resulting dynamics may lead to a shift of the average state in the system.
In the mean-field, we found that this shift only depends on the initial states of the nodes but, in general hypergraph topologies, this dependency is associated to the dominance of certain sub-graphs over others, due to
a complex interplay between the form of the interaction function, the topology of the cluster connection and the initialisation of the node states.

This work opens different research directions. First, we found an interesting relation between linear consensus dynamics on hypergraphs and the so-called motif Laplacian \cite{benson_higher_order_2016} proposed for community detection in higher-order networks. This relations suggests a way to generalise dynamics-based community detection methods for hypergraphs \cite{rosvall2009map,lambiotte2014random}. 
Another important extension to this work would be to go beyond three-body interactions and thus consider hyperedges of any cardinality. A natural way to achieve this goal would be, for each hyper-edge, to weight the influence of a node $j$ on node $i$ by its distance to the average opinion in the hyperedge (excluding node $i$). For three-body interactions, as considered here, this construction reduces to a scaled 3CM as
\begin{align*}
	\dot{x}_i &= \left|x_j - \frac{1}{2}(x_j + x_k)\right|\left[(x_j - x_i) + (x_k-x_i)\right] \\ 	
	&= \frac{1}{2}\left|x_j - x_k\right|\left[(x_j - x_i) + (x_k-x_i)\right].
\end{align*}

Other generalisations include the study of directed three-body interactions, which may include asymmetric roles to their constituents, but also of stochastic models of opinion dynamics like the majority rule \cite{krapivsky2003dynamics}, instead of the deterministic models considered here. Another limitation of our work comes from the very constrained hypergraph structures that we used, e.g. connected clusters, and the fact that the initialisation of the node states was aligned with  the topology. To better understand the dynamical properties of 3CM in a more practical setting, it would be essential to understand how, given a fixed hypergraph structure for instance obtained from empirical data, the initial configuration of the states would affect the asymptotic consensus. These findings could then be exploited in order to modify balances or imbalances in the dynamical system by seeding (i.e. changing the initial states) or eliminating components (i.e. changing the topology).
In a linear model of a two-body dynamical system, it is known that asymptotic properties are dominated by the structure, e.g. the mixing time is determined by the spectral gap. In the case of non-linear dynamics, the structure alone is not sufficient, and it becomes essential to understand how the states of the nodes affect notions such as convergence time and lead to an asymmetry between hypergraph components.

\begin{acknowledgments}
RL would like to thank Michael Schaub, Christian Bick, Ingo Scholtes and Martin Rosvall for inspiring discussions related to this manuscript. LN would like to thank the Hertie School for financial support.
AM is thankful for funding from the Oxford-Emirates Data Science Lab.
\end{acknowledgments}

\bibliographystyle{apsrev4-1}
\bibliography{references}

\end{document}

%% file: Pairwise_new.pdf_tex
\begingroup%
  \makeatletter%
  \providecommand\color[2][]{%
    \errmessage{(Inkscape) Color is used for the text in Inkscape, but the package 'color.sty' is not loaded}%
    \renewcommand\color[2][]{}%
  }%
  \providecommand\transparent[1]{%
    \errmessage{(Inkscape) Transparency is used (non-zero) for the text in Inkscape, but the package 'transparent.sty' is not loaded}%
    \renewcommand\transparent[1]{}%
  }%
  \providecommand\rotatebox[2]{#2}%
  \newcommand*\fsize{\dimexpr\f@size pt\relax}%
  \newcommand*\lineheight[1]{\fontsize{\fsize}{#1\fsize}\selectfont}%
  \ifx\svgwidth\undefined%
    \setlength{\unitlength}{876.93501802bp}%
    \ifx\svgscale\undefined%
      \relax%
    \else%
      \setlength{\unitlength}{\unitlength * \real{\svgscale}}%
    \fi%
  \else%
    \setlength{\unitlength}{\svgwidth}%
  \fi%
  \global\let\svgwidth\undefined%
  \global\let\svgscale\undefined%
  \makeatother%
  \begin{picture}(1,0.43984363)%
    \lineheight{1}%
    \setlength\tabcolsep{0pt}%
    \put(0,0){\includegraphics[width=\unitlength,page=1]{Pairwise_new.pdf}}%
    \put(0.48365909,0.29307475){\color[rgb]{0,0,0}\makebox(0,0)[lt]{\begin{minipage}{0.12531796\unitlength}\raggedright $i$\end{minipage}}}%
    \put(0.53893268,0.36178383){\color[rgb]{0,0,0}\makebox(0,0)[lt]{\begin{minipage}{0.46643853\unitlength}\raggedright $\color{red}f_{ij}$\end{minipage}}}%
    \put(0.73379202,0.39094677){\color[rgb]{0,0,0}\makebox(0,0)[lt]{\begin{minipage}{0.10332579\unitlength}\raggedright $j$\end{minipage}}}%
    \put(0.71913058,0.10690106){\color[rgb]{0,0,0}\makebox(0,0)[lt]{\begin{minipage}{0.12043082\unitlength}\raggedright $k$\end{minipage}}}%
    \put(0,0){\includegraphics[width=\unitlength,page=2]{Pairwise_new.pdf}}%
    \put(0.53753986,0.1020815){\color[rgb]{0,0,0}\makebox(0,0)[lt]{\begin{minipage}{0.48598717\unitlength}\raggedright $\color{red}f_{ik}$\end{minipage}}}%
  \end{picture}%
\endgroup%

%% file: Threeway.pdf_tex
\begingroup%
  \makeatletter%
  \providecommand\color[2][]{%
    \errmessage{(Inkscape) Color is used for the text in Inkscape, but the package 'color.sty' is not loaded}%
    \renewcommand\color[2][]{}%
  }%
  \providecommand\transparent[1]{%
    \errmessage{(Inkscape) Transparency is used (non-zero) for the text in Inkscape, but the package 'transparent.sty' is not loaded}%
    \renewcommand\transparent[1]{}%
  }%
  \providecommand\rotatebox[2]{#2}%
  \newcommand*\fsize{\dimexpr\f@size pt\relax}%
  \newcommand*\lineheight[1]{\fontsize{\fsize}{#1\fsize}\selectfont}%
  \ifx\svgwidth\undefined%
    \setlength{\unitlength}{987.36301318bp}%
    \ifx\svgscale\undefined%
      \relax%
    \else%
      \setlength{\unitlength}{\unitlength * \real{\svgscale}}%
    \fi%
  \else%
    \setlength{\unitlength}{\svgwidth}%
  \fi%
  \global\let\svgwidth\undefined%
  \global\let\svgscale\undefined%
  \makeatother%
  \begin{picture}(1,0.39058084)%
    \lineheight{1}%
    \setlength\tabcolsep{0pt}%
    \put(0,0){\includegraphics[width=\unitlength,page=1]{Threeway.pdf}}%
    \put(0.42279202,0.26144477){\color[rgb]{0,0,0}\makebox(0,0)[lt]{\begin{minipage}{0.10913195\unitlength}\raggedright $i$\end{minipage}}}%
    \put(0,0){\includegraphics[width=\unitlength,page=2]{Threeway.pdf}}%
    \put(0.64524436,0.34718373){\color[rgb]{0,0,0}\makebox(0,0)[lt]{\begin{minipage}{0.08742912\unitlength}\raggedright $j$\end{minipage}}}%
    \put(0.45417601,0.32386793){\color[rgb]{0,0,0}\makebox(0,0)[lt]{\begin{minipage}{0.5618508\unitlength}\raggedright $\color{blue}f_i^{\{jk\}}$\end{minipage}}}%
    \put(0,0){\includegraphics[width=\unitlength,page=3]{Threeway.pdf}}%
    \put(0.63164904,0.08540614){\color[rgb]{0,0,0}\makebox(0,0)[lt]{\begin{minipage}{0.10696168\unitlength}\raggedright $k$\end{minipage}}}%
  \end{picture}%
\endgroup%

%% file: influence_joint.pdf_tex
\begingroup%
  \makeatletter%
  \providecommand\color[2][]{%
    \errmessage{(Inkscape) Color is used for the text in Inkscape, but the package 'color.sty' is not loaded}%
    \renewcommand\color[2][]{}%
  }%
  \providecommand\transparent[1]{%
    \errmessage{(Inkscape) Transparency is used (non-zero) for the text in Inkscape, but the package 'transparent.sty' is not loaded}%
    \renewcommand\transparent[1]{}%
  }%
  \providecommand\rotatebox[2]{#2}%
  \newcommand*\fsize{\dimexpr\f@size pt\relax}%
  \newcommand*\lineheight[1]{\fontsize{\fsize}{#1\fsize}\selectfont}%
  \ifx\svgwidth\undefined%
    \setlength{\unitlength}{527.14283861bp}%
    \ifx\svgscale\undefined%
      \relax%
    \else%
      \setlength{\unitlength}{\unitlength * \real{\svgscale}}%
    \fi%
  \else%
    \setlength{\unitlength}{\svgwidth}%
  \fi%
  \global\let\svgwidth\undefined%
  \global\let\svgscale\undefined%
  \makeatother%
  \begin{picture}(1,1.28710906)%
    \lineheight{1}%
    \setlength\tabcolsep{0pt}%
    \put(0,0){\includegraphics[width=\unitlength,page=1]{influence_joint.pdf}}%
    \put(0.09094286,0.97743477){\color[rgb]{0,0,0}\makebox(0,0)[lt]{\begin{minipage}{0.20847425\unitlength}\raggedright $i$\end{minipage}}}%
    \put(0.73876193,1.29878757){\color[rgb]{0,0,0}\makebox(0,0)[lt]{\begin{minipage}{0.1718889\unitlength}\raggedright $j$\end{minipage}}}%
    \put(0.71843674,0.41975574){\color[rgb]{0,0,0}\makebox(0,0)[lt]{\begin{minipage}{0.20034419\unitlength}\raggedright $k$\end{minipage}}}%
    \put(0,0){\includegraphics[width=\unitlength,page=2]{influence_joint.pdf}}%
    \put(0.09780754,0.11130414){\color[rgb]{0,0,0}\makebox(0,0)[lt]{\begin{minipage}{0.20034419\unitlength}\raggedright $p$\end{minipage}}}%
  \end{picture}%
\endgroup%